\documentclass[prd,reprint,nofootinbib,nodate,superscriptaddress,amsmath,amssymb,aps,preprintnumbers]{revtex4-2}

\usepackage[utf8]{inputenc}
\usepackage{amsmath,amsthm,amsfonts,amssymb,amscd}
\usepackage{multirow,booktabs}
\usepackage[table]{xcolor}
\usepackage{fullpage}
\usepackage{lastpage}
\usepackage{enumitem}
\usepackage{fancyhdr}
\usepackage{mathrsfs}
\usepackage{wrapfig}
\usepackage{setspace}
\usepackage{calc}
\usepackage{cancel}
\usepackage{graphicx,bm,appendix}
\usepackage{hyperref}
\newcommand{\dd}{\mathrm{d}}
\usepackage{setspace}

\pdfoutput=1

\usepackage[utf8]{inputenc}

\usepackage{xcolor}

\usepackage{graphicx}
\usepackage{dcolumn}
\usepackage{bm}

\newcommand{\mA}{\mathcal{A}}

\newcommand{\mM}{\mathcal{M}}

\newcommand{\mO}{\mathcal{O}}

\newcommand{\dr}{(\rho-\rho_c)}
\newcommand{\sq}{ \sqrt{C_4 \dr}}

\begin{document}

\author{Mario Flory}
\email{mflory@th.if.uj.edu.pl}
\affiliation{Institute of Theoretical Physics, Jagiellonian University, {\L}ojasiewicza 11, 
30-348 Krak{\'o}w, Poland}
\affiliation{Instituto de F{\'i}sica Te{\'o}rica UAM-CSIC, c/ Nicol{\'a}s Cabrera 13-15, 28049, Madrid, Spain}

\author{Sebastian Grieninger}
\email{sebastian.grieninger@stonybrook.edu}
\affiliation{Center for Nuclear Theory, Department of Physics and Astronomy, Stony Brook University, New York 11794-3800, USA}
\affiliation{Instituto de F{\'i}sica Te{\'o}rica UAM-CSIC, c/ Nicol{\'a}s Cabrera 13-15, 28049, Madrid, Spain}
\affiliation{Departamento de F{\'i}sica Te{\'o}rica, Universidad Aut{\'o}noma de Madrid, Campus de Cantoblanco, 28049 Madrid, Spain}

\author{Sergio Morales-Tejera}
\email{sergio.moralest@uam.es}
\affiliation{Instituto de F{\'i}sica Te{\'o}rica UAM-CSIC, c/ Nicol{\'a}s Cabrera 13-15, 28049, Madrid, Spain}
\affiliation{Departamento de F{\'i}sica Te{\'o}rica, Universidad Aut{\'o}noma de Madrid, Campus de Cantoblanco, 28049 Madrid, Spain}

\title{\ \ \\
Critical and near-critical relaxation of holographic superfluids}
\preprint{IFT-UAM/CSIC-22-108} 
\date{\today}

\begin{abstract}
We investigate the relaxation of holographic superfluids after quenches, when the end state is either tuned to be exactly at the critical point, or very close to it. By solving the bulk equations of motion numerically, we demonstrate that in the former case the system exhibits a power law falloff as well as an emergent discrete scale invariance. The latter case is in the regime dominated by critical slowing down, and we show that there is an intermediate time-range before the onset of late time exponential falloff, where the system behaves similarly to the critical point with its power law falloff. We further postulate a phenomenological Gross–Pitaevskii-like equation (corresponding to Model F of Hohenberg\,\&\,Halperin) that is able to make quantitative predictions for the behavior of the holographic superfluid after near-critical quenches into the superfluid and normal phase. Intriguingly, all parameters of our phenomenological equation which describes the non-linear time evolution may be fixed with information from the static equilibrium solutions and linear response theory.   
\end{abstract}

\maketitle

\section{Introduction} 
\noindent

The (complex) Ginzburg-Landau and Gross-Pitaevskii equations are among the simplest yet 
most important nonlinear phenomenological models in mathematical physics, and can describe the surprisingly rich macroscopic behavior of various complex systems from nonlinear waves to non-equilibrium phenomena in  superconductors and superfluids 
\cite{RevModPhys.74.99,tsuneto_1998}. Relaxation of homogeneous superfluids near the second order phase transition is usually governed by the derivative of the free energy. However, exactly at the critical point, the derivative of the free energy vanishes and equilibration is totally governed by nonlinearities (inaccessible within linear response).

In experiments, it is impossible to engineer the system to relax exactly to the critical point. Hence, extending the analysis to \textit{nearly} critical quenches is necessary. Moreover, fluctuations of the order parameter are usually neglected in conventional hydrodynamics since they are captured by massive modes. However, close to the critical point those modes become light and have to be taken into account. For example, order parameter fluctuations were crucial in recent experiments on strange metals~\cite{Seibold_2021,Arpaia_2021}.

In the past decades, the AdS/CFT correspondence \cite{Maldacena:1997re,Gubser:1998bc,Witten:1998qj} has emerged as a new tool to investigate out-of equilibrium physics, especially in the strongly coupled regime \cite{Zaanen:2015oix}. Indeed, so-called holographic superconductors have been a steady target of research since their inception in \cite{Gubser:2008px,Hartnoll:2008vx,Hartnoll:2008kx,Herzog:2008he}. Despite the established nomenclature, it is commonly understood that these systems might more correctly be termed holographic superfluids, as the gauge-symmetry in the bulk should translate into a global symmetry according to common AdS/CFT wisdom 
\footnote{However, the Maxwell equations on the boundary can be imposed, leading to a genuine holographic superconductor \cite{Domenech:2010nf}.}. 
Another interesting perspective suggested in \cite{Maeda:2010br} is to understand large gauge transformations in the bulk as giving rise to transformations that are "background-gauge invariance" transformations in the boundary theory, affecting the sources imposed there without being tied to a dynamical photon. This perspective will be especially influential for our work.

There are many open problems in the use of holographic methods to model out-of equilibrium dynamics of superfluids, and this is an active research area. For instance, there is no qualitative understanding of the emergence of discrete scale invariance after exactly critical quenches observed in \cite{Erdmenger:2016msd}. More generally, it is not clear whether or to what degree the dynamics of a holographic superconductor can be captured by an effective theory formulated in the boundary language \cite{PhysRevLett.127.101601,Yan:2022jfc}. We seek to make progress in this direction by studying critical and near-critical quenches in a holographic superconductor both from a bulk and a boundary perspective. This leads us to postulate a phenomenological Gross–Pitaevskii-like equation incorporating background-gauge invariance that is able to make quantitative predictions for the dynamics after critical and near-critical
quenches in the cases we study. 

\newpage

\section{Bulk results}
\noindent We consider the model of a holographic superconductor in AdS$_4$/CFT$_3$ \cite{Hartnoll:2008vx,Hartnoll:2008kx,Herzog:2008he,Gubser:2008px,Herzog:2009xv}: 
\begin{widetext}
\begin{equation}
\label{eq:action}
S\,=\,S_\text{grav}+
\frac{1}{2\kappa^2}\int_{\mathcal{M
}} \dd^{4}x\, \sqrt{-g}\left[
-\frac{1}{4\,q^2}F_{\mu\nu}F^{\mu\nu}-|D\varphi|^2-m^2|\varphi|^2\right]+S_\text{nf}\,.
\end{equation}
\end{widetext}

The fields include the metric $g_{\mu \nu}$, a $U(1)$ gauge field $A_{\mu}$ ($F=dA$) and a complex scalar field $\varphi$ of mass $m^2=-2$ charged under the gauge field with $D_\mu\equiv\partial_\mu-i\,q\,A_\mu$ and charge $q=+1$. Neglecting backreaction, we fix the metric
\begin{align}\label{eq:metric}
&\dd s^2=\frac{1}{u^2}\left[-f(u)\,\dd t^2-2\, \dd t \dd u+ \dd x^2+\dd y^2\right],
\end{align}
with $f(u)=1-u^3$, setting $u_h=1$ and $L=1$. We measure all physical quantities in terms of the fixed temperature $\bar T=4\pi T/3$ where $T=|f'(1)|/(4\pi)$. 
To simplify the notation we assume that all quantities have been divided by the appropriate power of $\bar T$ and hence are dimensionless~\footnote{ The dimensionless ratios are $\mu/{\bar T}, \,\rho/{\bar T}^2,\, \langle\mathcal O\rangle/{\bar T}^2,\, t\,\bar T, \,C_2 {\bar T}^3,\,C_3{\bar T},\,C_4/{\bar T}^2,\,C_5{\bar T}^3$\label{foot1}}. $S_{\text{nf}}$ is an external source which we use to perform a quench in the charge density. 
The equations of motion (eoms) are solved by a fully pseudo-spectral code 
as previously employed in~\cite{Ammon:2016fru}.

The near boundary expansions of the bulk matter fields read
\begin{align}
\varphi&\sim 
\langle \mathcal{O} \rangle u^2+\ldots,\ 
A_t
\sim \mA_t-\rho u+\ldots,
\end{align}
where we set the source of the scalar field to zero in order to engineer spontaneous symmetry breaking
and $2\kappa^2\langle\mO\rangle \equiv \Psi(t)=\phi(t)e^{i\psi(t)}$ is the complex expectation value of the dual operator~\cite{Arean:2021tks}. At static equilibrium we identify $\mA_t=\mu$ where $\mu$ is the chemical potential. The subleading component $\rho(t)$ is the charge density. There is a second order phase transition to a phase with non-zero condensate at $\rho=\rho_c\approx 4.06371$. We perform quenches of the system by giving $\rho$ a step-function-like time dependence. See appendix \ref{app::numerical} for more details.

We want to study how holographic superconductors relax to their equilibrium state after a quench when the final state is close to the critical point. Unlike in the Kibble-Zurek mechanism~\cite{Chesler:2014gya,Sonner:2014tca,Bhaseen:2012gg,delCampo:2021rak,Li:2021jqk}, we are interested in quenches with initial state in the ordered phase, i.e.~the quenches we are studying do not cross the critical point at any finite rate.   
For a \textit{near}-critical end-state this relaxation will be characterized at late times by an exponential falloff, where the half-life time diverges as the end-state is taken towards the critical point.  This is known as \textit{critical slowing down} \cite{RevModPhys.49.435}, see \cite{Maeda:2008hn,Maeda:2009wv,Natsuume:2010vb,Ewerz:2014tua} for holographic works. 
But what if the end-state is tuned to lie exactly at the critical point? In this case, the exponential falloff is replaced by a power law falloff \cite{Janssen,PhysRevB.96.054303,PhysRevB.81.012303}. This makes sense, as a power law falls off slower than \textit{any} exponential.
From the holographic bulk perspective this may be seen as the system trying to balance the condensate floating above the horizon exactly, hence the slow decay. The condensate is affected by electrostatic forces driving it into the bulk and gravitational forces trying to pull it into the horizon~\cite{Gubser:2008wv}.

\begin{figure}[htbp]
\includegraphics[width=0.49\textwidth]
{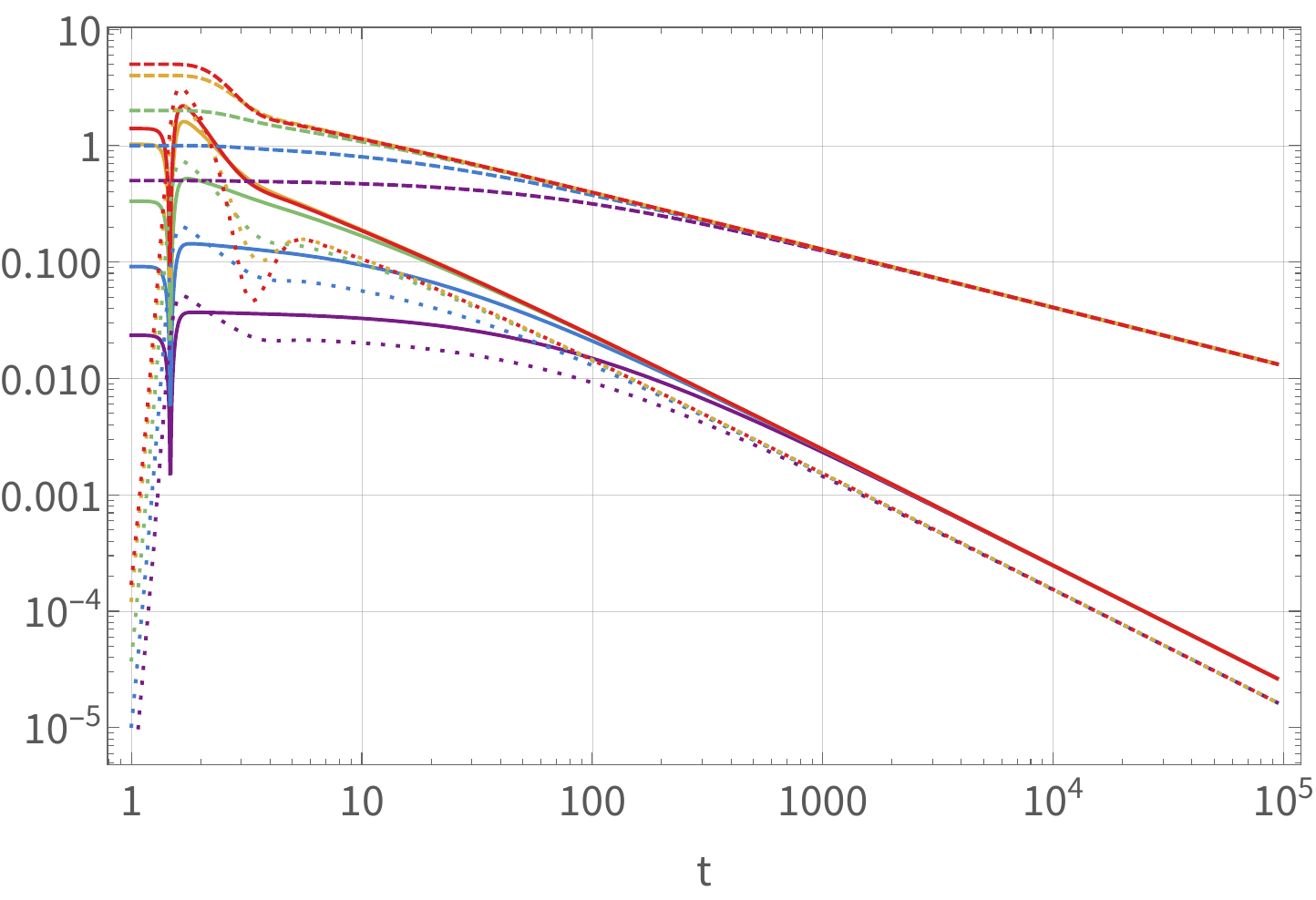}
	\caption{
	Numerical results for $|\mA_t(t)-\rho_c|$ (solid lines), $|\langle\mO\rangle|\equiv\phi(t)$ (dashed lines), and 
	$|\dot{\psi}(t)|$ (dotted lines) for multiple exactly critical quenches starting from $\rho_{\text{initial}}=\{4.124,4.302,4.951,6.981,8.184\}$ (purple, blue, green, orange, red). We find a universal power-law late time behavior with $\phi\propto 1/\sqrt{t}$ and $|\mA_t-\rho_c|,|\dot{\psi}|\propto 1/t$.
	}
	\label{fig::critical}
\end{figure}

Figure \ref{fig::critical} shows representative numerical results. We clearly observe late time power-law falloffs in $\phi(t)$, $|\dot{\psi}(t)|$, and $|\mA_t(t)-\rho_c|$ that are universal, i.e.~independent on the initial state or other details of the quench. 
At late times, where $\rho(t)$ is constant, the bulk eoms are not explicitly dependent on $t$, and so for any solution $y(t)$, $y(t+\delta t)$ will also be a solution for any $\delta t$. We hence make the ansatz 
\begin{align}
    \phi(t)&=A (t+\delta t)^\alpha
    \label{model1}
    \\
    \dot{\psi}(t)-(\mA_t(t)-\rho_c)&=B (t+\delta t)^\gamma
    \label{model2}
\end{align}
which we fit to our numerical curves at late times ($t\geq 5\times 10^4$). We only consider the combination in equation \eqref{model2} because it is gauge invariant under background-gauge transformations \cite{Maeda:2010br}. 
With high consistency between the quenches plotted in figure \ref{fig::critical}, we obtain
\begin{align}
\begin{tabular}{lllllll}
$A$ & $\approx$ & 4.07 
\ \  \ \  \ \  & $\alpha$ & $\approx$ & $-0.50$ 
\\
$B$ & $\approx$ & 0.93
\ \  &$\gamma$ & $\approx$ & $-1.00$ 
\end{tabular}
\label{fitted}
\end{align}
with only the value of $\delta t$ varying significantly from quench to quench.

The observed behavior $|\dot{\psi}|\propto1/t$ indicates $\psi(t)\propto\log{t}$, i.e.~oscillations of the real- and imaginary part of  $|\langle\mO\rangle|$ that are periodic on a logarithmic time axis. This signifies the presence of a \textit{discrete scale invariance}, in contrast to the continuous scale invariance inherent to ordinary power laws \cite{SORNETTE1998239}. 
Such discrete scale invariance has also previously been observed numerically in the formation of black holes through the collapse of charged scalar fields \cite{PhysRevD.51.4198} and
after critical quenches in a holographic Kondo model in \cite{Erdmenger:2016msd}, as well as in other holographic setups \cite{Liu:2009dm,Faulkner:2009wj,Hartnoll:2015rza,Balasubramanian:2013ux,Flory:2017mal,Ammon:2018wzb}. 

\section{Boundary model}
\noindent From a boundary point of view, the phenomenology of superconductors is described by the (complex) Ginzburg–Landau equation, while for superfluids with their global symmetry breaking the Gross–Pitaevskii equation takes a similar role  \cite{RevModPhys.74.99,tsuneto_1998}. Consequently, there has been some recent activity in 
\cite{PhysRevLett.127.101601,Yan:2022jfc} trying to fit parameters of Gross–Pitaevskii equations in order to model aspects of the non-equilibrium behavior of holographic superfluids.

Unlike the quenches that we study here, \cite{PhysRevLett.127.101601,Yan:2022jfc} investigated inhomogeneous setups where space-derivatives are non-zero which increases the complexity of the problem significantly. 
In addition, we explicitly study behavior near or even exactly at the critical point which means that our results should be ideally suited for such a phenomenological description, since e.g.~the Ginzburg-Landau equation is usually seen as a series expansion around vanishing order-parameter, where higher order terms in the free energy are dropped.


\begin{widetext}
We now postulate the phenomenological equation
\begin{align}
   & \left[\partial_t - i C_1  \left(\mA_t(t) -\rho + C_5 |\Psi(t)|^2\right) \right]\Psi(t)
\equiv -(C_2+ i C_3) \left[|\Psi(t)|^2 - C_4(\rho-\rho_c) \right]\Psi(t),
    \label{eq2}
\end{align}
\end{widetext}
where again $\Psi=\phi e^{i\psi}$,  $\rho,\rho_c,\psi,\mA_t,C_i\in\mathbb{R}$, $\phi>0$, and we neglect any terms including spatial derivatives or higher orders of $\Psi$ or $\rho-\rho_c$. The parameter $\rho$ is assumed to be constant in time, and $C_1$ is the charge of the complex field which in our case is $+1$. As we show in appendix \ref{app::ModelF}, this equation is exactly what is expected from Model F in the classification of Hohenberg and Halperin \cite{RevModPhys.49.435}, and the matching between holographic superconductors and this Model F has indeed been investigated also in a number of other recent papers \cite{Donos:2022qao,Donos:2023ibv,Bu:2024oyz}.   

The right hand side of \eqref{eq2} is similar to the variation of the free energy that would appear in the Ginzburg-Landau equation, multiplied with a complex prefactor which is inspired by the dissipative Gross-Pitaevski equations used in 
\cite{RevModPhys.74.99,PhysRevLett.127.101601,Yan:2022jfc}. 
The left hand side is essentially a gauge-covariant time-derivative plus some extra terms whose relevance will become clear shortly. The gauge covariance is necessary in order to respect the background gauge-invariance 
which arises as a consequence of large gauge transformations which do not fall off towards the boundary and hence change the boundary values of the bulk fields such as $\mA_t$ \cite{Maeda:2010br}. 

Because \eqref{eq2} contains complex factors, we can split it into a real and imaginary part (after dividing by $e^{i \psi}$ on both sides). Ignoring the trivial case $\phi=0$, the real part of this equation only depends on $\phi(t)$ and has the exact solution 
\begin{align}
    \phi(t)=\sqrt{\frac{C_4 (\rho-\rho_c )}{1-\left(1-\frac{C_4 (\rho-\rho_c )}{\phi_0^2}\right) e^{-2 C_2 C_4 t (\rho-\rho_c )}}}
    \label{phi}
\end{align}
with $\phi(0)=\phi_0$. This can be plugged into the imaginary part of \eqref{eq2} in order to obtain an algebraic equation for the gauge-invariant expression $\dot{\psi}-C_1\mA_{t}$. 
Obtaining a unique solution for both $\psi(t)$ \textit{and} $\mA_t(t)$ requires some kind of gauge fixing condition.

As a simple consistency check, looking for non-trivial ($\phi\neq0$) static solutions yields
\begin{align}
    \phi&=\sqrt{C_4 \dr}
    \label{phiEq}
    \\
     \mA_t&=\rho-C_5 \phi^2
    =\text{$\rho $}-C_4 C_5 (\rho -\rho_c)
    \label{AtEq}   
\end{align}
when $C_4 \dr>0$, i.e.~we observe the expected formation of the superfluid phase, and the deviation of the chemical potential $\mu=\mA_t$ from its value $\mu=\rho$ in the normal phase. 

In the non-equilibrium case, the late time exponential falloff of \eqref{phi} and $\dot{\psi}-C_1\mA_{t}$ 
depends on whether we are in the superfluid phase ($\rho>\rho_c$) 
or in the normal phase ($\rho<\rho_c$). 
The crucial insight is that we can determine the phenomenological parameters $C_i$ by comparing our results so far to the properties of the static solution in the superfluid phase and the late time quasinormal mode like falloff towards this static solution after a non-critical quench. As explained in appendix \ref{app::numerical}, this allows to numerically fix the parameters (normalized to $\bar T$) of the model to 
\begin{align}
\begin{tabular}{lllllll}
$C_2$ & $\approx$ & 0.03018\ \  \ \  \ \  & $C_3$ & $\approx$ & 0.09308 \\
$C_4$ & $\approx$ & 4.09192\ \  &$C_5$ & $\approx$ & 0.14967 
\end{tabular}\ .
\label{Ci}
\end{align}
With this, the model \eqref{eq2} is now able to make predictions for the behavior of the system at earlier times after non-critical quenches (via \eqref{phi}) as well as the exactly critical quenches where at late times $\rho = \rho_c$. In the latter case, \eqref{phi} simplifies to 
\begin{align}
    \phi(t)=\frac{1}{\sqrt{2 C_2 t+\frac{1}{\phi_0^2}}}\approx\frac{4.07}{t^{1/2}}+...
    \label{phicrit}
\end{align}
and additionally we find
\begin{align}
    \dot{\psi}-C_1(\mA_{t}-\rho_c)=\frac{C_1 C_5+C_3}{2 C_2 t+\frac{1}{\phi_0^2}}\approx\frac{0.94}{t}+...\ .
        \label{psicrit}
\end{align}
Although these solutions are small at late times, it is important to note that they depend on the non-linear features of equation \eqref{eq2} and could not be obtained through a linearized ansatz. 
Comparing the predictions \eqref{phicrit} and \eqref{psicrit} with equations \eqref{model1}, \eqref{model2}, and \eqref{fitted}, we find excellent agreement. Equation \eqref{phicrit} matches the scaling solution derived in \cite{Janssen} for an "initial-slip exponent" $\theta=0$. This value was recently also observed for an AdS/QCD model \cite{Cao:2022mep}.

Let us now return to the issue of near-critical quenches, starting and ending near the critical point, but not exactly at it. At very late times, after such a quench the system will be characterized by an exponential falloff towards the new equilibrium, however equation \eqref{eq2} describes the behavior of the system already at much earlier times. 
A representative example of a near-critical quench is depicted in figure \ref{fig::nearcritical} (for a quench into the normal phase 
see appendix \ref{app::normal}). We can see that before the equilibrium is reached at very late times, there is a long intermediate stretch of time in which the condensate $\phi(t)$ appears to fall off in a power law-like manner.
In particular, after a quench that brings the system infinitesimally close to the critical point, the system will initially react as if it was relaxing exactly to the critical point, and only after what we call "handover-timescale"
$t_\text{ho}\sim \frac{1}{|\rho-\rho_c|}$
the system will notice that it is not at the critical point, and the power-law behavior gives way to an exponential falloff towards a small but finite condensate. Of course, $t_\text{ho}$ is identical to the relaxation timescale of the system close to the critical point.  
This phenomenological behavior is encapsulated in equation \eqref{phi}, as shown in appendix \ref{app::intermediate}.

As the parameters $C_i$ are given in \eqref{Ci} and the value $\rho$ is determined by the choice of quench, the only parameter that needs to be determined in order to compare our analytical prediction with the numerical result is $\phi_0$. 
As we can see in figure \ref{fig::nearcritical}, in contrast to $\dot{\psi}-\mA_t$, $\phi(t)$ does not change significantly during and immediately after the quench. Neither would a sudden change be predicted by \eqref{phi}. Hence, as we know the initial state before the quench exactly, we can simply set $\phi_0\equiv\phi(0)$, even though formally equations \eqref{eq2} and \eqref{phi} only become valid after the quench, when $\rho$ is constant.  As shown in figure \ref{fig::nearcritical}, this trick is sufficient to obtain a very good match between numerical data and analytical solution.

\begin{figure}[htbp]
\includegraphics[width=0.5\textwidth]{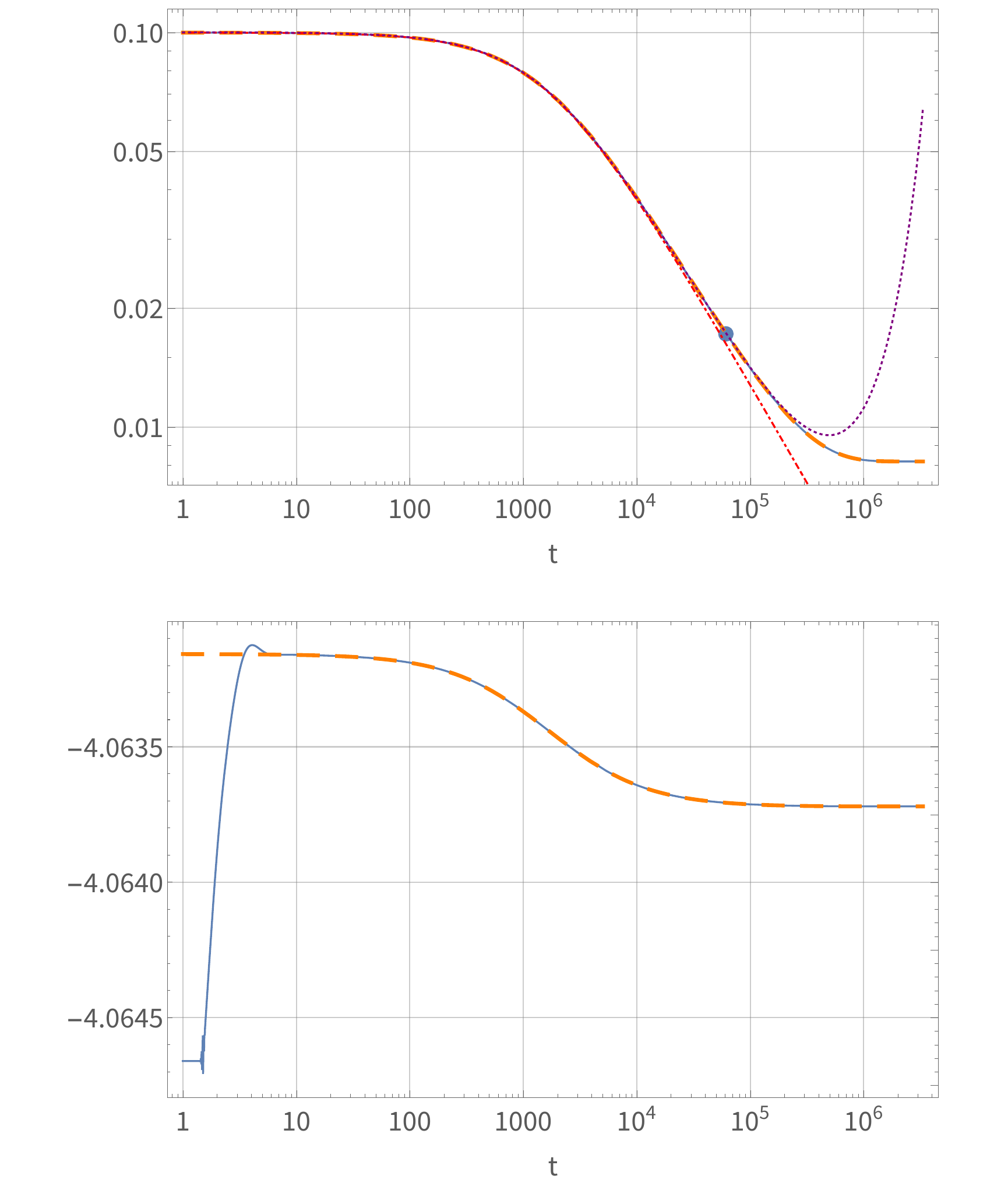}
	\caption{
 Numerical (solid blue) and analytical (dashed orange) results for a near-critical quench, with $\rho_{\text{initial}}=$4.06626 and $\rho_{\text{final}}=$4.06373. The top frame shows $\phi(t)$, the bottom frame shows $\dot{\psi}(t)-C_1 \mA_t(t)$. After the quench ($t\gtrsim10$), numerical and analytical curves agree very well.  
 The dot signifies the time scale $t_\text{ho}$, while the dash-dotted red line shows the critical solution \eqref{phicrit}. The dotted purple line shows the approximation $\phi(t)\approx(2C_2 t + 1/\phi_o^2)^{-1/2}e^{C_4(\rho-\rho_c)(2C_2t+1/\phi_0^2)/4}$ derived in appendix \ref{app::intermediate}. 
 }
	\label{fig::nearcritical}
\end{figure}

\section{Discussion} 
\noindent We studied the relaxation of a holographic superconductor close or exactly at the critical point. 
In the bulk, we used extensive numerical simulations over long ranges of time. 
We demonstrated that for critical quenches, in addition to the expected power law falloff of the modulus $|\langle \mO \rangle|$ of the order parameter, its complex phase undergoes rotations which are periodic on a logarithmic time axis, leading to a \textit{discrete scale invariance}. 
Furthermore, we observed that the power law falloff characteristic for the critical quenches is approximately observable even in the non-critical quenches for an intermediate time-period before the onset of the late-time exponential falloff at a handover-timescale $t_\text{ho}$.

On the field theory side, we postulated a phenomenological Gross–Pitaevskii-like equation for the system corresponding to Model F of \cite{RevModPhys.49.435} -- see appendix \ref{app::ModelF} for details. This equation can be solved analytically, and its parameters can be fixed by comparing to the numerical data on the late-time exponential falloff after non-critical quenches. In particular, we can fix the parameters of the equation with information about the static equilibrium states and linearized fluctuations about them (quasinormal modes (QNMs)) which are much easier to compute than the non-linear time evolution. The constants obtained from the QNM data in the superfluid and normal fluid phase match which is a non-trivial check of our suggested equation~\eqref{eq2}. In our study we also successfully applied and tested a novel concept about computing the amplitudes of QNM excitations developed in~\cite{Arean:2021tks}. 

Once the parameters are fixed, the phenomenological equation predicts with good accuracy the behavior both after exactly critical quenches, as well as after near-critical quenches for intermediate time-scales. Importantly, both in the bulk and on the boundary, our results are intimately tied to the non-linearities of the respective equations of motion, and could not be studied with a simple linearized ansatz (even though methods based on linearized equations have been shown to be surprisingly accurate in some circumstances \cite{Heller:2012km,Heller:2013oxa}). We hence established a non-trivial check that holographic superfluids do obey Model F of \cite{RevModPhys.49.435} even at the non-linear level, see appendix \ref{app::ModelF} for further details. Interestingly, the papers \cite{Mitman:2022qdl,Cheung:2022rbm} have recently commented on the limitations of linearized ansätze in the study of black hole ringdowns.   

Interesting future directions would be to include spacial dependence (similar to 
\cite{PhysRevLett.127.101601,Yan:2022jfc,Adams:2012pj,Sonner:2014tca}), 
to study systems with different symmetry groups (such as \cite{Cao:2022mep}), and to turn on backreaction on the metric in the bulk. Preliminary results indicate the possibility to generalize~\eqref{eq2} to a finite rate of superflow within the parameter regime of second order phase transitions up to the tricritical point. It might also be interesting to make contact between our results and the complementary approach of \cite{Bu:2021clf}.

\section{Acknowledgments}

\noindent We would like to thank Matteo Baggioli, Carlo Ewerz, Adrien Florio, Thomas Gasenzer, Karl Landsteiner, Hesam Soltanpanahi, and Derek Teaney for useful discussions. 

The work of MF (until the 31st of August 2022), SG and SM were supported through the grants CEX2020-001007-S and PGC2018-095976-B-C21, funded by MCIN/AEI/10.13039/501100011033 and by ERDF A way of making Europe. Since September 1st 2022, the work of MF was supported by the Polish National Science Centre (NCN) grant 2021/42/E/ST2/00234. SG was in part supported by the
``Juan de la Cierva- Formación'' program (grant number FJC2020-044057-I) funded by MCIN/AEI/10.13039/
501100011033 and NextGenerationEU/PRTR. SG is supported by the U.S. Department of Energy, Office of Science grants No. DE-FG-88ER40388. SM was further supported by an FPI-UAM predoctoral fellowship.

\appendix

\section{Numerical methods}
\label{app::numerical}

\noindent In this section, we give a brief overview of the numerical methods used to solve the partial differential equations numerically. 

Within numerical holography, pseudo-spectral methods are widely applied to find highly accurate solutions to boundary value problems in terms of elliptic partial differential equations or ordinary differential equations. However, for initial value problems of hyperbolic partial differential equations space and time are typically treated differently. The spatial dependence is usually discretized by means of a (pseudo)-spectral method which is combined with an explicit 4th order Runge-Kutta scheme or Adams-Bashforth method to evolve the solution in time~\cite{Chesler:2013lia}. Within a fully spectral scheme, we discretize time and space with (pseudo)-spectral methods yielding a highly implicit and accurate time evolver.

The basic idea of (pseudo)-spectral methods \cite{Boyd1989ChebyshevAF}~is that the unknown functions $u(x)$ which is the solution to the differential equation
\begin{equation}
    \mathcal L_x u(x)=g(x),
\end{equation}
where $\mathcal L_x$ is a differential operator, may be approximated by a finite number $N$ of basis polynomials $\phi_i(x)$
\begin{equation}
    u(x)\approx u_N(x)=\sum_{i=0}^{N-1}c_i\phi_i(x).
\end{equation}
To find the solution, we require that the residuum $R=\mathcal L_x u_N-g$ vanishes exactly on the chosen set of discrete grid points. Note that for the exact solution, the residuum vanishes identically. For a given choice of grid points and basis polynomials, the derivatives are replaced by discrete matrices acting on the whole domain.

Fully spectral algorithms have been employed within (asymptotically flat) numerical relativity in~\cite{Hennig:2008af,Hennig:2012zx,Petri2014ASM,PanossoMacedo:2014dnr,Hennig:2020rns} and in the context of holography in~\cite{Ammon:2016fru,Grieninger:2016xue}. Let us outline the recipe for our numerical algorithm (see also the appendix of~\cite{Grieninger:2016xue}).

\subsection{Numerical algorithm}
We are looking for a solution at time $t=t_\text{final}$ to the initial value problem at $t=t_\text{initial}$.
\begin{enumerate}
    \item At $t=t_\text{initial}$, we may obtain the initial configuration by solving the static set of ordinary differential equations subject to the boundary conditions $\varphi'(t_\text{initial},0)=0$ and $\varphi''(t_\text{initial},0)=|\langle\mathcal O_\text{initial}\rangle|$ and $A_t(t_\text{initial},1)=0$. As discretization of the radial direction $u\in[0,1]$, we chose Gauss-Lobatto grid points $u_j=\frac12\left(1+\cos(\pi j/N_u)\right)$, where $j\in[0,N_u-1]$.
    \item To evolve in time, we decompose the time interval $(t_\text{initial},t_\text{final}]=(t_\text{initial},t_1]\cup(t_1,t_2]\cup...\cup(t_p,t_\text{final}]$ in $p+1$ subintervals in the spirit of multi-domain decomposition. 
    Note that the different time intervals may have different sizes which we set by an adaptive step control depending on how much the solution changes on the interval.
    \item For a given initial solution, we introduce auxiliary functions \newline$h(t,u)=h_\text{in}(t=t_i,u)+(t-t_i)\,h_\text{aux}(t,u)$ on each subinterval $(t_i,t_j],\, t_j>t_i$.
    \item The radial coordinate is discretized by the Chebyshev-Lobatto (CL) grid and to discretize the time coordinate, we chose a right-sided Chebyshev-Radau (rCR) grid (for some generic time interval $t \in (t_i, t_j]$)
\begin{align}
  &  u_j = \frac12\left(1 + \cos\left(\frac{\pi j}{N_u}\right)\right),\\ &t_k = \frac12\left[ (t_j-t_i) + (t_j-t_i) \cos\left(\frac{2\pi k}{2N_t+1}\right)\right],
\end{align}
where $j=0,\ldots, N_u-1$, and $k = 0,\ldots, N_t$. Note that the right-sided Chebychev-Radau grid does not include the initial slice of the interval where we already know the solution.
\item Replace all derivatives by their discrete versions given by the derivative matrices $D$: $\partial_u\to D^\text{CL},\partial_t\to D^\text{rCR}$ and discretize the equations of motion on the square spanned by the discrete spectral coordinates and impose the desired boundary.
\item Solve the corresponding non-linear system with a Newton-Raphson method.
\item Use the solution on the final slice $t_j$ as the new initial solution in the next step.
\end{enumerate}
Typically, we use $N_u=40$ or $N_u=50$ in combination with $N_t=14$. We monitored that the constraint equation is satisfied better than $10^{-15}$ during the time evolution. Note that we additionally fix $A_t(t,1)=0$. The numerical algorithm is implemented in \textit{Mathematica}.

The numerical methods used to compute the initial configurations, background solutions and QNMs are the same codes as used in~\cite{Ammon:2021pyz}.

\subsection{Quench profiles}
\label{sec::quench_profiles}
For spontaneous $U(1)$ symmetry breaking, charge conservation imposes $\dot{\rho}(t)=0$ in the absence of external sources. However, our goal is to study quenches from the superfluid phase with $\rho=\rho_\text{initial}$ to the critical point $\rho=\rho_\text{crit}$ (or close to it for some $\rho=\rho_\text{final}$). Since we work in the probe limit, we have to introduce an external source in order to change $\rho$ to our desired final value. We could achieve this by breaking the $U(1)$ explicitly with a scalar source as done in~\cite{Bhaseen:2012gg}, or we consider some generic external source that changes $\rho$ directly by considering the null fluid (nf) current~\cite{Fernandez-Pendas:2019rkh}
\begin{equation}
   2\kappa^2\, J_{(\text{nf})}^u=\frac{\dot \rho}{\sqrt{-g}}.
\end{equation}
which may be achieved by coupling
\begin{equation}
    S_\text{nf} = \frac{1}{2\kappa^2}\int_{\mathcal{M}}\dd^4x \sqrt{-g}A_{\mu}J^{\mu}_\text{ext}\,,
\end{equation}
to the action. This leads to a covariantly conserved external current \begin{equation}
    J_\text{ext}^{\mu} = \dfrac{\dot{\rho}}{u^2}\delta^\mu_u\,.
\end{equation}
which allows us to change the electric charge at will.
Technically, the external source also drives the $T^{tt}$ component of the energy-momentum tensor. However, in the large $q$ expansion of the probe limit this contribution is subleading and we can neglect it similarly to the backreaction of the matter fields onto the geometry.

To quench our system, we chose to manipulate $\rho$ with the external source and quench it to its final value. Note that in the late time behavior, the external source is switched off and we verified that the late time behavior is independent of the quench profile. Concretely, we performed quenches with
\begin{equation}
   \rho(t)= \rho_\text{initial} +\frac12(\rho_\text{final} -\rho_\text{initial} )  (1 + \tanh[\Omega (t - t_s)]);
\end{equation}
where $\Omega=10$ is the rapidity and $t_s=1.5$ is the center of the quench.

\subsection{Determination of parameters}

The parameters $C_4,C_5$ of equation \eqref{eq2} can be determined by fitting equations \eqref{phiEq},\eqref{AtEq} to the behavior of the static holographic superconductor close to the critical point. One way to determine $C_2,C_3$ is to compare the predicted late time behavior after a non-critical quench into the superfluid phase to numerical data of the non-linear time evolution. With the parameters fixed in this way, \eqref{eq2} then allows to make genuine predictions for both exactly critical quenches, and for the behavior of non-critical quenches at early and intermediate times. 

We find that for $t\gg 1$, quenches in the superfluid phase will relax as
\begin{widetext}
\begin{align}
   & \phi(t)=\sqrt{C_4} \sqrt{\rho -\rho_c}+\frac{\sqrt{C_4}}{2} \sqrt{\rho -\rho_c} \left(1-\frac{C_4 (\rho-\rho_c )}{\phi_0^2}\right) e^{-2 C_2 C_4 t (\rho -\rho_c)}+...
   \label{latetime1}
   \end{align}
   \begin{align}
&\dot{\psi}(t)  -C_1\mA_t(t)  =-C_1 \rho 
+C_1 C_4 C_5 (\rho -\rho_c)
-\frac{C_4  (C_1 C_5-C_3) }{\phi_0^2}\left(C_4 (\rho -\rho_c)-\phi_0^2\right)(\rho -\rho_c) e^{-2 C_2 C_4 t (\rho-\rho_c )}+...
   \label{latetime2}
\end{align}
while in the normal phase we would find 
\begin{align}
   & \phi(t)=
   \sqrt{\frac{C_4(\rho_c - \rho)}{1-\frac{C_4 (\rho-\rho_c )}{\phi_0^2}}}
   e^{- C_2 C_4 t (\rho_c -\rho)}+...
   \label{latetime1uncon}
   \\
 &  \dot{\psi}(t)  -C_1\mA_t(t)  =-C_1 \rho +C_3 C_4 (\rho-\rho_c )
+\frac{C_4 \phi_0^2 (\rho -\rho_c) (C_1 C_5-C_3)}{C_4 (\rho -\rho_c)-\phi_0^2}
e^{-2 C_2 C_4 t (\rho_c -\rho)}+... \ .
   \label{latetime2uncon}
\end{align}
\end{widetext}

Clearly, $C_2$ can be determined by fitting the half-life time of the predicted late time exponential falloff in the superfluid phase to the numerical data of the non-linear time evolution. The amplitude of this exponential falloff itself is of course dependent on $\phi_0$ for each of these functions, however the ratio between the amplitudes 
\begin{align}
    \frac{\text{Amplitude}_\phi}{\text{Amplitude}_{\dot{\psi}(t)  -C_1\mA_t(t)}}=\frac{1}{2\sq}\frac{1}{C_1 C_5- C_3}\label{eq:ampl}
\end{align}
is independent of $\phi_0$ and can be used to obtain a unique value of $C_3$.

However, it is \textbf{not} necessary to perform the non-linear time evolution to fit the constants \eqref{Ci}. In the following, we explain how all of them may be obtained within linear response theory and from the static solutions in the context of holography. The two constants $C_4$ and $C_5$ may simply be obtained by constructing the static solutions near the phase transition and fit the condensate and chemical potential, respectively, to the deviation of $\rho$ from its critical value $\rho_c$ according to eqs.~\eqref{phiEq} and~\eqref{AtEq}. The constant $C_2$ may be obtained from the QNMs at zero wavevector in the superfluid phase~\eqref{latetime1} or in the normal phase~\eqref{latetime1uncon}.
To probe the QNMs we consider linearized fluctuations of a complex scalar $\delta\bar\Psi=\delta\Psi e^{-i\omega t}$ and the gauge field $\delta\bar a_t=\delta a_t e^{-i\omega t}$. Note that we may decompose the scalar fluctuations about a state with zero background phase according to $(\langle\mathcal O_\text{eq}\rangle+\langle\delta\mathcal O\rangle)e^{i\delta\psi}=(\langle\mathcal O_\text{eq}\rangle+\langle\delta\mathcal O\rangle)(1+i\delta\psi)=(\langle\mathcal O_\text{eq}\rangle+\text{Re}(\delta\Psi)+i\,\text{Im}(\delta\bar\Psi)$, with $\text{Re}(\delta\Psi)=\langle\delta\mathcal O\rangle)$ and $\text{Im}(\delta\bar\Psi)=\langle\mathcal O_\text{eq}\rangle\,\delta\psi$.

Let us focus on the normal phase first and consider linearized solutions about the static normal phase solution with $\langle\mathcal O\rangle=0,\, \rho<\rho_c$. The corresponding QNM responsible for the relaxation to equilibrium is the pair of massive scalar modes. Close to the phase transition the QNMs in the normal phase behave (to lowest order in $\rho-\rho_c$) like
\begin{equation}
   \omega_\pm= -(\pm 0.38087-0.12348i)\,(\rho-\rho_c)\label{QNMvaluncond}
\end{equation}
According to eq.~\eqref{latetime1uncon}, we can read of the constant $C_2$ (since we already know $C_4$ from the static solution) from the imaginary part of the QNM~\eqref{QNMvaluncond} leading to the value indicated in~\eqref{Ci}. Similarly, the real part determines the constant $C_3=-\text{Re}(\omega_+)/C_4$ as may be seen from equation \eqref{latetime2uncon}. Since the QNM comes as a pair, the sign is seemingly not determined. However, it is possible to reconstruct which sign belongs to fluctuations of $\delta\Psi$ and $\delta\bar\Psi$. In order to determine the QNMs we solve the fluctuation equations as generalized eigenvalue problem of the form $(\bm{A}\omega-\bm{B})\,\bm{x}=0$, where $\bm{A}$ and $\bm{B}$ are differential operators of a non-hermitian Sturm-Liouville problem (see~\cite{Arean:2021tks} for more details). Usually, only the eigenvalues $\omega$ are of interest since they correspond to the QNM frequencies. However, it is also possible to examine the eigenvector $\bm{x}$ corresponding to the eigenvalue $\omega$. In our case, we observe that fluctuations with $\omega=\omega_+$ are carried by $\bm{x}=\{\delta\Psi,0\}$ while $\omega=\omega_-$ is carried by $\bm{x}=\{0,\delta\bar\Psi\}$.

As an independent, non-trivial check of our proposed equation, we now compute the constants $C_2$ and $C_3$ from the QNMs in the superfluid phase. To compute $C_3$, we need information about the relative amplitudes of the fluctuations supporting the QNM responsible for equilibration. Only recently, the authors of~\cite{Arean:2021tks} suggested a method to compute the relative contributions of boundary operators to a certain QNM excitation from the aforementioned eigenvectors. Here, we want to dissect the so called ``amplitude'' or Higgs mode. At zero wave vector, this pseudo-diffusive mode is driving the system to equilibrium in the superfluid phase~\cite{Amado:2009ts,Amado:2013xya}. More recently, the dynamics of this mode was discussed in terms of a linearized bulk analysis in~\cite{Donos:2022xfd}. Close to equilibrium, we find for the QNM frequency to lowest order in $\rho-\rho_c$ (and at zero wave vector)
\begin{equation}
    \omega_\text{Ampl}= -0.2469\, i \,(\rho-\rho_c).
\end{equation}
According to eq.~\eqref{latetime1}, we may extract $C_2$ from this information leading to the same numerical value as computed in the normal phase.
Employing the techniques developed in~\cite{Arean:2021tks}, we can extract the expectation values of the operators carrying this QNM excitation from the corresponding eigenvector. Intriguingly, we find that the gauge fluctuations have expectation value zero and the mode is solely carried by the scalar fluctuations. Close to the critical point we thus find to lowest order in $\rho-\rho_c$ (and at zero wave vector) that
    \begin{align}
    \frac{\text{Amplitude}_{\langle\delta\mathcal O\rangle}}{\text{Amplitude}_{\langle\delta\dot\psi\rangle-\langle\delta a_t\rangle}}=\frac{17.67}{ 2\,\sqrt{C_4(\rho-\rho_c)}}.\label{eq:amplQNM}
\end{align}
Once $C_5$ is fixed from the background data, equations \eqref{eq:ampl} and \eqref{eq:amplQNM} determine the value of $C_3$ in accordance with our data from the normal phase. Note that this is a non-trivial and independent check of the numerical values we obtained for $C_2$ and $C_3$ thus confirming the prediction of our suggested model.

\section{Analysis of intermediate time behavior}
\label{app::intermediate}

\noindent We will now give an analysis of the behavior at intermediate time scales predicted by the solution \eqref{phi} as well as the corresponding solution

\begin{widetext}
\begin{align}
\dot{\psi}-C_1\mA_{t}
= -C_1 \rho - C_3 C_4 (\rho-\rho_c)+\frac{C_4(C_3+C_1 C_5)(\rho-\rho_c)}{1-e^{-2 C_2 C_4t(\rho-\rho_c)}\left(1-\frac{C_4(\rho-\rho_c)}{\phi_0^2}\right)}.
\label{A}
\end{align}
\end{widetext}

First of all, we notice that if the final state is in the condensed phase, then $\frac{C_4(\rho-\rho_c)}{\phi_0^2}=\frac{\phi(t\rightarrow\infty)^2}{\phi(0)^2}$. As we have been interested exclusively in quenches that lead to a decay of the condensate, we will assume $\frac{C_4(\rho-\rho_c)}{\phi_0^2}<1$. Hence the bracket depending on $\phi_0$ in \eqref{phi} and \eqref{A} is positive and can be absorbed in the exponent as a shift $t\rightarrow t+t_0$ in the time coordinate. As we would like to ignore such shifts, we will from now on take the limit $\phi_0\rightarrow\infty$. This limit is technically un-physical, as equation \eqref{eq2} is only expected to be reliable for small values of $\phi$, however it simplifies equations \eqref{phi} and \eqref{A} and their analysis considerably. The results of this analysis should then also hold for realistic settings up to shifts on the time axis.

We already discussed the very late time behavior in equations \eqref{latetime1} and \eqref{latetime2}, seeing an exponential falloff towards equilibrium at $t\gg1$. 
Now, we turn our attention to earlier times. For this, we define a map $\mM(y(t))\equiv \frac{\dot{y}(t)\times t}{y(t)}$. This has the benefit that it easily allows us to analyze and distinguish the qualitative behavior of functions, as e.g.~
\begin{align}
\mM(A t^a)&=a,
\\
\mM(A e^{at})&=a t,
\\
\mM(A t^b e^{at})&=b+a t.
\end{align}
For our solutions \eqref{phi} and \eqref{A}, we  find 
\begin{widetext}
\begin{align}
\mM(\phi(t))&=-\frac{1}{2}+\frac{1}{2}C_2 C_4 \dr t + ...,
\label{MofPhi}
\end{align}
\begin{align}
\mM(\dot{\psi}-C_1(\mA_{t}-\rho_c))&=-1 + \frac{C_2(-C_3 C_4+C_1(-2+C_4C_5))}{C_3+C_1C_5}\dr t+... \ .
\end{align}
\end{widetext}

\begin{figure}[htbp]
\includegraphics[width=0.5\textwidth]
{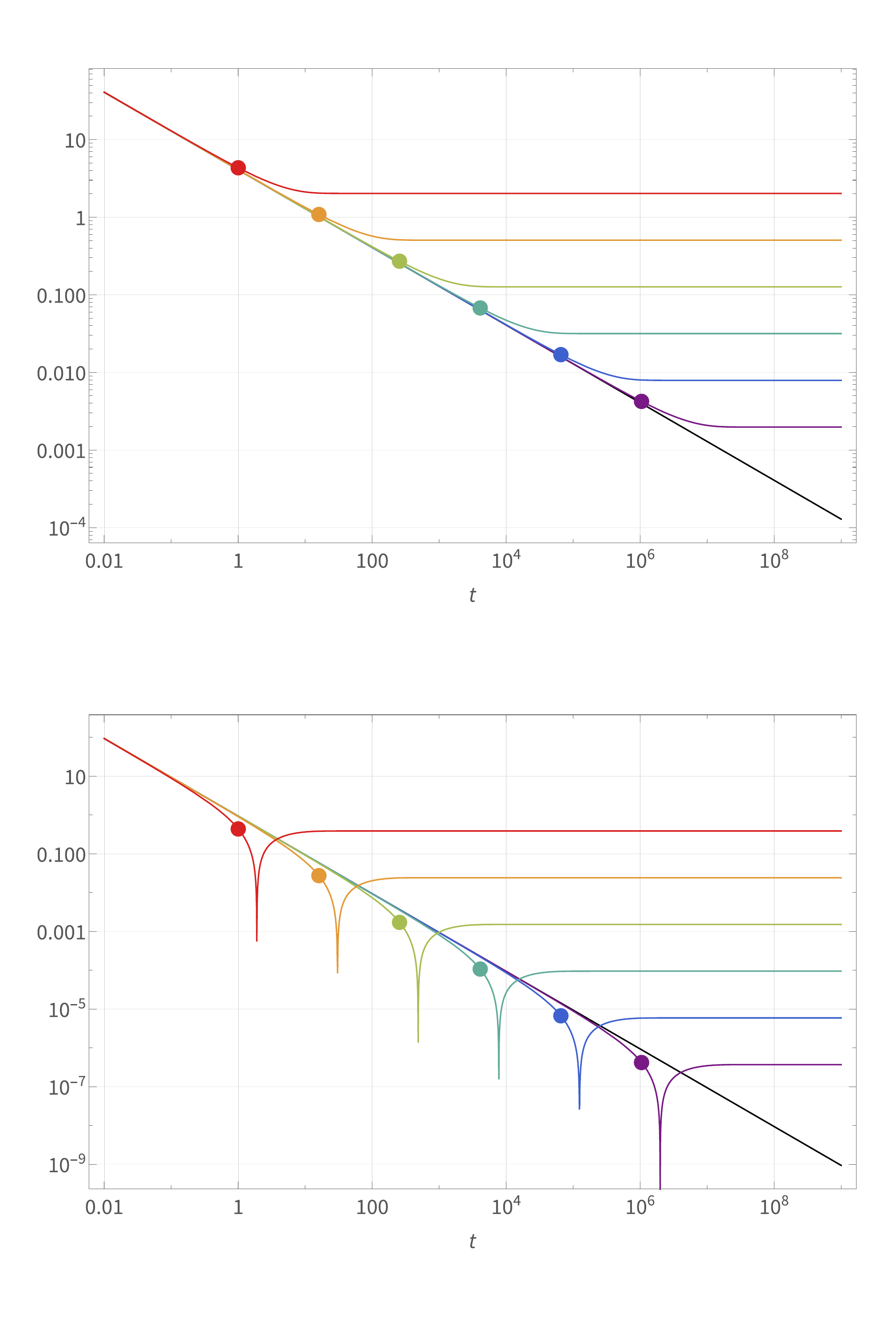}
	\caption{
 The top frame shows $\phi(t)$ (using \eqref{phi}) and the bottom frame shows $|\dot{\psi}-C_1(\mA_{t}-\rho_c)|$(using \eqref{A}) for $\phi_0\rightarrow\infty$ and $\rho=\rho_c+2^a$ with $a$ varying from $a=-20$ (purple) to $a=0$ (red) in steps of $4$. The black lines represent the exactly critical solutions $\rho=\rho_c$. The dots placed on each curve signify the handover timescale $t_\text{ho}$, which clearly describes well in an order of magnitude manner until what timescale the solutions are well approximated by the critical solutions \eqref{phicrit} and \eqref{psicrit}.
 }
	\label{fig::ho}
\end{figure}

Assuming all coefficients $C_i$ to be roughly of order 1, this demonstrates that the solutions for non-critical quenches exhibit the same kind of power law behavior as the critical quenches until $\dr t$ is of order 1, which establishes the handover timescale $t_\text{ho}$. See figure \ref{fig::ho} for an illustration. More precisely, we could have said based on \eqref{MofPhi} that for early times $\phi(t)$ can be approximated as
\begin{align}
\phi(t)\propto t^{-1/2} e^{\frac{1}{2}C_2 C_4 \dr t},
\label{betterapprox}
\end{align}
however for $t\ll t_\text{ho}$ the exponential function will deviate from 1 only slightly.

\begin{figure}[htbp]
\includegraphics[width=0.5\textwidth]{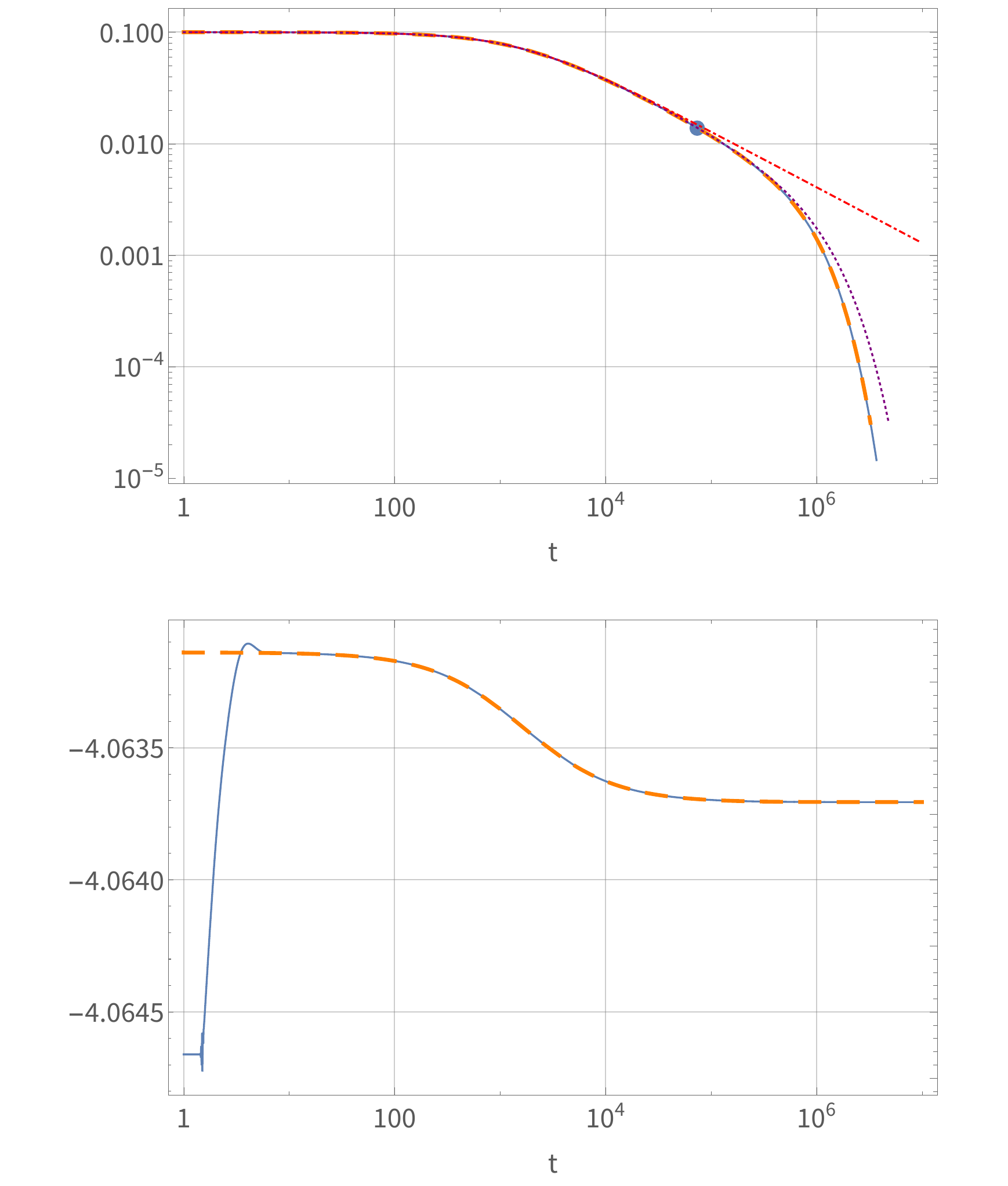}
	\caption{
 Comparison between numerical (solid blue) and analytical (dashed orange) results for a near-critical quench, with $\rho_{\text{initial}}=4.06626$ and $\rho_{\text{final}}=4.06370$. The top frame shows $\phi(t)$, while the bottom frame shows $\dot{\psi}(t)-C_1 \mA_t(t)$. For the times after the quench ($t\gtrsim10$), numerical and analytical curves agree very well. 
 The dot signifies the handover-time scale $t_\text{ho}$, while the dash-dotted red line shows the critical solution \eqref{phicrit}. The dotted purple line shows the approximation $\phi(t)\approx(2C_2 t + 1/\phi_o^2)^{-1/2}e^{C_4(\rho-\rho_c)(2C_2t+1/\phi_0^2)/4}$ derived in \eqref{betterapprox} (where we had assumed $\phi_0\rightarrow\infty$). 
 }
	\label{fig::nearcriticalnormal}
\end{figure}

\section{Near-critical quenches in normal phase}
\label{app::normal}

Here, we briefly demonstrate that the phenomenological equation~\eqref{eq2} can also describe near-critical quenches into the normal phase. In figure~\ref{fig::nearcriticalnormal}, we show the normal phase analogue of the quench shown in figure~\ref{fig::nearcritical} of the main text, i.e.~we pick a $\rho_\text{final}$ near the critical point but in the normal phase. As evident from figure~\ref{fig::nearcriticalnormal}, the intermediate and late time behavior are in perfect agreement with the numerical solution from holography.

\section{Comparing to Model F}
\label{app::ModelF}

In the absence of external fields or noise terms, Model F as defined in \cite{RevModPhys.49.435} is given by

\begin{equation}\label{eq:F_1}
   \partial_t \Psi= -2 \Gamma_0 \dfrac{\delta F_0}{\delta \Psi^*}- i g_0 \Psi \dfrac{\delta F_0}{\delta m},
\end{equation}

\begin{equation}\label{eq:F_2}
    \partial_t m = \lambda_0 \nabla^2 \dfrac{\partial F_0}{\partial m}+ 2 g_0 \textrm{Im}\left(\Psi^* \dfrac{\delta F_0}{\delta \Psi^*}\right),
\end{equation}
where $\Psi =\phi e^{i\psi}$ is the complex valued order parameter and $m$ is the conserved charge density. The model allows for $\Gamma_0 \in \mathbb{C}$ while $g_0\,,\lambda_0 \in \mathbb{R}$. The functional $F_0$ is defined through \cite{RevModPhys.49.435}: 
\begin{align}
    F_0 = &\int d^d x \left(  \frac{1}{2}\tilde{r}_0 |\Psi|^2 + \dfrac{1}{2}|\nabla \Psi|^2 + \tilde{u}_0|\Psi|^4\right.\\ &\left. + \dfrac{1}{2C_0} m^2 + \gamma_0 m |\Psi|^2\right)\,.\nonumber
\end{align}
In the homogeneous case, equation \eqref{eq:F_1} reduces to
\begin{align}\label{eq:F_4}
    &\left( \partial_t - i\frac{g_0}{C_0}\left(-m - \gamma_0 C_0 |\Psi|^2  \right) \right)\Psi =\\ & -4 \tilde{u_0}\Gamma_0 \left(|\Psi|^2 + \dfrac{\tilde{r}_0 + 2\gamma_0 m}{4 \tilde{u}_0}\right)\Psi\,.
    \nonumber
\end{align}
To make contact with our model \eqref{eq2}, we identify the complex order parameters ($\Psi$ in both formulations) and the conserved quantity 
 \begin{equation}
     m=\rho.
 \end{equation}
Moreover, for the matching we adopt the gauge choice
 \begin{align}
\mA_t=0. 
 \end{align}
This gauge choice corresponds to the Josephson relation used in \cite{RevModPhys.49.435}.
Comparing the parameters in the models \eqref{eq:F_1} and \eqref{eq2} results in the following relations:
\begin{align}
C_1 &= \frac{g_0}{C_0}
\\
C_2+iC_3&=4\tilde{u}_0 \Gamma_0
    \\
    C_4&=-\frac{\gamma_0}{2\tilde{u}_0}
    \\
    C_5&=- C_0\gamma_0
    \\
    \rho_c&=-\frac{\tilde{r}_0}{2\gamma_0}.
\end{align}
Note that in the homogeneous case where all spatial derivatives vanish, equation \eqref{eq:F_2} trivially simplifies to $\partial_t m =0$ as the right-hand side is the imaginary part of a manifestly real expression. Our model \eqref{eq2} (where we had implicitly assumed charge conservation, $\partial_t\rho=0$, from immediately after the quench onwards, see appendix \ref{sec::quench_profiles}) is hence equivalent to the predictions of Model F. 
Shortly after the first draft of this manuscript had been posted on the arXiv, explicit comparisons between the holographic superfluid and Model F also appeared in \cite{Donos:2022qao,Donos:2023ibv,Bu:2024oyz}. 
Specifically, in \cite{Donos:2022qao} a matching of the holographic superfluid to Model F was put forward to linear order and explicit expression for the parameters of Model F were found in terms of horizon data. This is somewhat complementary to our approach; firstly, we tried to model the holographic superconductor (and fix the parameters $C_i$ of our model) from a purely boundary perspective, as if the numerical data on the time-dependence of boundary observables which we calculated from holography were experimental data given to us. Secondly, we numerically solve the full non-linear bulk equations (which can describe dynamics far from equilibrium) and not small linearized perturbations about an equilibrium solution. 
\newpage
\bibliography{references}
\bibliographystyle{bibstyl}
\end{document}